\documentstyle[aps,twocolumn]{revtex}
\begin{document}
\preprint{USACH-FM-0105 }
\title{The Landau problem and noncommutative quantum mechanics}
\author{J. Gamboa$^1$\thanks{E-mail: jgamboa@lauca.usach.cl}, M. Loewe
$^2$\thanks{Email: mloewe@fis.puc.cl},
F. M\'endez$^1$\thanks{E-mail: fmendez@lauca.usach.cl} and J. C. Rojas$^3$
\thanks{E-
mail: rojas@sonia.ecm.ub.es}}
\address{$^1$Departamento de F\'{\i}sica, Universidad de Santiago de Chile,
Casilla 307, Santiago 2, Chile \\
$^2$Facultad de F\'{\i}sica, Pontificia Universidad Cat\'olica de Chile, Casilla 306,
Santiago 22, Chile
\\
$^3$Departament ECM, Facultat de Fisica, Universitat de Barcelona and Institut D'Altes Energies,
\\
Diagonal 647, E-08028, Barcelona, Spain }
\maketitle
\begin{abstract}
The conditions under which noncommutative quantum mechanics and the Landau problem are equivalent
theories is explored. If the potential in noncommutative quantum mechanics is chosen as
$V= \Omega \,\aleph$ with $\aleph$ defined in the text,  then for the value
${\tilde \theta} = 0.22 \times 10^{-11} \,cm^2$ (that measures the noncommutative effects of the space), the
Landau problem and noncommutative quantum mechanics are equivalent theories in the lowest Landau
level. For other systems one can find differents values for  ${\tilde \theta}$ and, therefore, the possible
bounds for ${\tilde \theta}$ should be searched in a physical independent scenario. This last fact could
explain the differents bounds for $\tilde \theta$ found in the literature.
\end{abstract}
\pacs{03.65.-w, 03.65.Db}
\narrowtext

The presence of magnetic fields in the string effective action suggest a noncommutative structure for the
spacetime. This fact and the possibility of compactifying string theory via the noncommutative torus
\cite{string} have stimulated an important amount of work during the last three years\cite{varios}.

A noncommutative space is an intriguing and revolutionary possibility that could have important
consequences in our conception of the quantum structure of nature. A noncommutative space is related to
the fundamental new commutation relation
\begin{equation}
[x,y] \sim \alpha  \theta, \,\,\,\, \label{1}
\end{equation}
where we will define $\alpha \theta$ as ${\tilde \theta}$ that is a parameter with dimensions
$(\mbox{lenght})^2$.

In a previous paper\cite{glr} the noncommutative quantum mechanics for a two dimensional central field was
studied and an explicit realization of the Seiberg-Witten map, at quantum mechanical level, was found.  In
this letter we would like to argue that the usual Landau problem also can be understood in terms of
noncommutative quantum mechanics. Additionally, we will show that the experimental value for the magnetic
field in the QHE\cite{vk}  implies
\begin{equation}
{\tilde \theta} \sim   0.22\times 10^{-11} \,\, cm^2, \label{pa}
\end{equation}
{\it i.e.} the characteristic magnetic length can be derived from noncommutative quantum mechanics.

Let us start  with the Moyal product for potential term in the Schr\"odinger equation in a noncommutative
plane
\begin{equation}
V({\bf x}) \star  \psi ({\bf x})  = V({\bf x} -  \frac{1}{2}{\tilde {\bf p}})\psi ({\bf x}), \label{bopp}
\end{equation}
where $\star$ denotes the Moyal product and $\tilde p _{i_k}=\theta^{i_kj_k} p_{j_k}$
$\theta _{ij} = \theta \epsilon _{ij}$.

Then for a two dimensional central field\cite{glr} $V({\vert {\bf x}\vert}^2)$, (\ref{bopp}) yields
\begin{equation}
V({\vert {\bf x}\vert}^2) \star \psi ({\bf x})  = V({\hat \aleph})\psi ({\bf x}), \label{c1}
\end{equation}
where the aleph ($\aleph$) operator is defined as
 \begin{eqnarray}
 {\hat \aleph} &=& \frac{{\theta}^2}{4} p_x^2 + x^2 + \frac{{\theta}^2} {4}
 p_y^2 + y^2  -{ \theta} L_z
 \label{c2}
\\
&=& {\hat H}_{HO} - {\theta}{\hat L}_z. \label{c3}
\end{eqnarray}

In the last expression ${\hat H}_{HO}$ is the hamiltonian for a two dimensional harmonic oscillator with mass
$2/{\theta}^2$, frequency $\omega = {\theta}$ and $L_z$ is the z-component of the angular
momentum
defined
as $L_z = x p_y - y p_x$.

From (\ref{c1}) one see that $\aleph$ has $(\mbox{lenght})^2$ and, furthermore, the dimensions of
${\theta}$
are $\mbox{time/mass}$. This last fact imply that in (\ref{1}) one must choice $\alpha = \hbar$ in (\ref{1}), {\it
i.e.}
\begin{equation}
{\tilde \theta} = \hbar\, \theta, \label{eff}
\end{equation}

 Thus, from (\ref{1}) one can think that ${\tilde \theta}$ effectively measure the noncommutative effects of the
 space.

The eigenstates and the eigenvalues of the $\aleph$ operator were computed in \cite{glr} and the result is
\begin{equation}
{\hat \aleph} \vert jm> ={ \theta} \,[\,2j + 1- 2m] \vert jm>, \label{c4}
\end{equation}
with the selection rules
\begin{eqnarray}
j&=& 0, \frac{1}{2}, 1, \frac{3}{2}, ...\nonumber
\\
m&=& j, j-1, j-2, ..., -j . \label{selec}
\end{eqnarray}

Thus, the hamiltonian for noncommutative quantum mechanics in a central field becomes
\begin{equation}
{\hat H} = \frac{1}{2M} p^2 + V({\hat \aleph}).\label{c5}
\end{equation}

If we choose the potential
\begin{equation}
V({\hat \aleph}) = \Omega \aleph, \label{5}
\end{equation}
being $\Omega$ an appropriate constant, then the hamiltonian (\ref{c5}) can be written as
\begin{equation}
H = (\frac{1}{2M} + \frac{\Omega\, {\theta}^2}{4}) (p_x^2+ p_y^2) + \Omega\, (x^2 + y^2) - \Omega\, {
\theta} L_z.
\label{land1}
\end{equation}

Our next step is to consider the two-dimensional Landau problem in the symmetric gauge, whose
Hamiltonian for a particle with mass $\mu$ is given by
\begin{equation}
{\hat H}_{\text {Landau}} = \frac{1}{2\mu} ( p_x^2 + p_y^2)+ \frac{e^2H^2_0}{8\mu} (x^2+y^2)  -
\frac{
eH_0}{2 \mu}
L_z. \label{a1}
\end{equation}

 Now we note from  (\ref{land1}) and (\ref{a1}) that these  two problems are equivalent. This equivalence
 means that the magnetic field must be strong enough in order to confine  the particles in the plane $x-y$ .
 After this identification, the following relations are satisfied
\begin{eqnarray}
 \frac{1}{2M} + \frac{\Omega\, {\theta}^2}{4} &=& \frac{1}{2\mu}
 \\
\frac{e^2H^2_0}{8\mu}&=&\Omega, \label{para}
\\
\Omega \,{\theta} &=& \frac{eH_0}{2 \mu} \label{theta}
\end{eqnarray}

These equations are consistent if and only if $M=\infty$\cite{bigatti}. Furthermore, the Hamiltonians
(\ref{land1}) and (\ref{a1})  describes the same noncommutative system in the lowest Landau level, {\it i.e.} in
the strong regime of the magnetic field.

Using (\ref{para}) and (\ref{theta}) one find that
\begin{equation}
{\tilde \theta} = \frac{4\hbar}{e H_0} , \label{est}
\end{equation}

In the QHE experiments\cite{vk}, the magnetic fields are  typically about 12T. In this way we get
\begin{equation}
{\tilde \theta} = 0.22\times 10^{-11}\, cm^2 \label{est1}
\end{equation}

For this value of ${\tilde\theta}$ one cannot distinguish between noncommutative quantum
mechanics\cite{glr} and the usual Landau problem. This route was explored by Bellisard\cite{belli} and
other authors\cite{jackiw,algunas} using different points of view (for other bounds for $\theta$ see \cite{kos}
and references therein). Qualitatively one could also observe noncommutative effects in the Aharanov-Bohm
experiments as was proposed in \cite{glr2}.

Finally we would like to point out the following: 1) the magnetic field in equation (\ref{est}) cannot be
arbitrarily small due to the identification we have done assuming the existence of the lowest Landau  level
and 2) the value for ${\tilde \theta}$ is strongly  model dependent. Therefore, if spacetime is
physically realized as a noncommutative structure, then the bounds  for ${\tilde \theta}$ should have a
universal character and should be independent of a particular physical scenario. One possibility could be to
explore Lorentz violation invariance as in\cite{kos2}.
\acknowledgments
We would like to thank Prof. V.O. Rivelles and V. A. Kostelecky by comments on the manuscript. This work
was partially supported by FONDECYT-Chile under grant numbers 1010596, 1010976 and 3000005.

\end{document}